**Spatially mixed moiré excitons in two-dimensional van der Waals superlattices**

Yanhao Tang[1], Jie Gu[1], Song Liu[2], Kenji Watanabe[3], Takashi Taniguchi[3], James Hone[2], Kin Fai Mak[1,4,5]*, Jie Shan[1,4,5]*

[1]School of Applied and Engineering Physics, Cornell University, Ithaca, NY, USA
[2]Department of Mechanical Engineering, Columbia University, New York, NY, USA
[3]National Institute for Materials Science, 1-1 Namiki, 305-0044 Tsukuba, Japan
[4]Laboratory of Atomic and Solid State Physics, Cornell University, Ithaca, NY, USA
[5]Kavli Institute at Cornell for Nanoscale Science, Ithaca, NY, USA
*Email: kinfai.mak@cornell.edu; jie.shan@cornell.edu

**Moiré superlattices open an unprecedented opportunity for tailoring interactions between quantum particles [1–11] and their coupling to electromagnetic fields [12–17]. Strong superlattice potential generates moiré minibands of excitons [16,17] -- bound pairs of electrons and holes that reside either in a single layer (intralayer excitons) or two separate layers (interlayer excitons). The twist-angle-controlled interlayer hybridization of carriers can also mix the two types of excitons to combine the strengths of both [13,18,19]. Here, we report a direct observation of spatially mixed moiré excitons in angle-aligned $WSe_2/WS_2$ and $MoSe_2/WS_2$ superlattices by optical reflectance spectroscopy. The strongly interacting interlayer and intralayer moiré excitons in $WSe_2/WS_2$ manifest energy level anticrossing and oscillator strength redistribution under a vertical electric field. We also observe doping-dependent exciton miniband renormalization and mixing near half filling of the first electron miniband of $WS_2$. Our findings have significant implications for emerging correlated states in two-dimensional semiconductors, such as exciton condensates [20] and Bose-Hubbard models [21], and optoelectronic applications of these materials.**

Interlayer excitons in two-dimensional (2D) semiconductor heterostructures possess an out-of-plane electric dipole, which enables electric-field-tuning of resonance energies by the quantum-confined Stark effect [22–25]. The long lifetime of interlayer excitons also facilitates long-range exciton transport [26,27] and population buildup for the realization of exciton condensation [20]. However, unlike (bright) intralayer excitons that interact strongly with light [28], interlayer excitons have negligible oscillator strength because of the spatially indirect nature. This significantly limits their fundamental studies and applications. To gain oscillator strength, interlayer excitons can mix with intralayer excitons through interlayer hybridization of electrons or holes. Energy level anticrossing would be a hallmark signature of strong coupling. Spatially mixed excitons have been intensively searched in semiconductor heterostructures [29]. Recent experiments on 2D transition metal dichalcogenide (TMD) moiré superlattices attributed new spectral features observed in reflectance contrast to spatially mixed excitons [13,18,19,30,31]. The superlattice potential provides extra Bragg reflection for quasi-momentum conservation, opens new optical transition paths and facilitates hybridization between intralayer and interlayer excitons. While these studies show the twist-angle-dependent exciton-band structure, direct evidence of spatially mixed excitons and their continuous tuning, particularly in the presence of strong moiré potential, remain elusive.

Here we report direct observation of spatially mixed moiré excitons in angle-aligned WSe$_2$/WS$_2$ and MoSe$_2$/WS$_2$ heterobilayers. They exhibit the characteristics of both an interlayer exciton (large out-of-plane electric dipole) and an intralayer exciton (appreciable oscillator strength). The spatially mixed moiré excitons are formed via spin-conserving resonant tunneling of carriers between the layers. They are both electric-field-tunable and doping-tunable. In particular, we observe energy level anticrossing in WSe$_2$/WS$_2$ heterobilayers with strong moiré potential. In MoSe$_2$/WS$_2$ heterobilayers, we observe multiple families of spatially mixed moiré excitons with distinct electric dipoles. The different families are likely associated with exciton localization in different sites of the moiré superlattice. Finally, the characteristic doping dependence of the uncoupled exciton miniband edges at half filling of the first electronic miniband reveals rich manybody interaction effects induced by moiré superlattices in WSe$_2$/WS$_2$ heterobilayers.

Monolayer TMD semiconductors possess a direct energy gap at the K and K' valleys of the Brillouin zone [28] (Fig. 1b, c). At these valleys the large spin-orbit coupling pins the electronic spin in the out-of-plane direction and spin-splits the bands. The splitting is ~ 10's and 100's meV, respectively, at the conduction and valence band edges. The spin of each state flips when the momentum changes from K to K' so that time reversal symmetry is preserved. We choose heterobilayers of WSe$_2$/WS$_2$ and MoSe$_2$/WS$_2$, in which closely aligned bands are present to enhance interlayer hybridization.

We fabricate TMD heterobilayers by stacking two monolayers with a controlled twist angle between their crystal lattices. We employ a dual-gate device (Fig. 1a) to apply vertical electric field and doping to the heterobilayers independently. The two nearly symmetric gates are made of hexagonal boron nitride (hBN) of ~ 20 nm in thickness and few-layer graphite. The heterobilayers are grounded through another piece of few-layer graphite. When equal gate voltages of the same sign are applied to the gates, only doping is introduced to the heterobilayers. When equal gate voltages of opposite sign are applied, only vertical electric field $E_{appl}$ is applied. Similar results have been observed in multiple devices in this work. (See Methods for details on device fabrication and calibration of the doping level and electric field.)

Figure 1d - f are the reflectance contrast spectra $R$ of WSe$_2$/WS$_2$ heterobilayers with twist angles close to 20°, 60° and 0° as a function of gate voltage $V_G$. The reflectance contrast was measured as a ratio, $R \equiv (I' - I)/I$, between the reflected light intensity from the sample ($I'$) and from the substrate ($I$). Here the gate voltage is applied symmetrically to the two gates, i.e. varying the doping density with $E_{appl} = 0$. The three horizontal dashed lines correspond to filling factor $\nu = 0$, 1 and 2 for electrons in WS$_2$ ('$\nu = 2$' corresponds to full filling of the first electron moiré minband). WSe$_2$/WS$_2$ heterobilayers are known to have a type-II band alignment [3,11] (Fig. 1b, c). The lower valence band of WSe$_2$ is closely aligned with the upper valence band of WS$_2$. The lowest-energy spatially mixed excitons (dashed double-arrowed line), if present, are expected to occur near the fundamental exciton resonance of WS$_2$ (solid double-arrowed line). We focus on a spectral window centered on the fundamental exciton of WS$_2$ (the most prominent feature around 2 eV).

Our result is consistent with recent reports [3,12]. In particular, the WS$_2$ fundamental exciton weakens with electron doping. Effects of the moiré superlattice are observed only in angle-aligned samples (0° and 60°). They manifest as multiple minibands of intralayer excitons, which exhibit semi-periodic modulations at integer filling factors. In the 60° sample we also observe a new spectral feature (iX) slightly below the WS$_2$ fundamental exciton resonance (X$_1$).

We examine the vertical electric-field response of the new spectral feature iX to understand its origin. Figure 2a-c shows results for the 60° sample with $\nu$ close to 0, 1 and 2, respectively. The result for the 0° sample is included in Supplementary Fig. 1 for comparison. No discernible electric-field dependence is observed for moiré excitons in the 0° sample. This is consistent with the fact that intralayer excitons do not have out-of-plane electric dipoles. In contrast, electric-field dependence is observed for all features in the 60° sample. In addition to iX, two intralayer moiré excitons (X$_1$ and X$_2$) of WS$_2$ are also visible. The fundamental intralayer moiré exciton (X$_1$) carries the largest oscillator strength. When iX is far from the X$_1$ and X$_2$ excitons, it disperses linearly with field and has small oscillator strength. In this regime, X$_1$ and X$_2$ show little dispersion with field. When iX approaches the other exciton resonances, energy level anticrossing is observed and iX rapidly gains oscillator strength.

The linear vertical electric-field response and negligible oscillator strength of iX (when it is decoupled from other excitons) support that iX is an interlayer exciton [22–24]. It arises from transitions from the lower valence band of WSe$_2$ to the conduction band of the same spin in WS$_2$ (Fig. 1b). Energy level anticrossing and oscillator strength redistribution between the interlayer and intralayer excitons signify their strong coupling. The phenomenon is observed only in samples with twist angles close to 60°. In this configuration, extra Bragg reflection from the superlattice potential helps conserve quasi-momentum and facilitates strong coupling. In addition, the closely aligned valence bands in the two layers have the same spin, which makes resonant hole tunneling possible when the bands are tuned into resonance by electric field. This is in contrast to the 0° sample (Fig. 1c), in which resonant tunneling is spin-forbidden and no spatially mixed excitons can be seen. We thus conclude that the spatially mixed excitons in WSe$_2$/WS$_2$ heterobilayers are formed through spin-preserving resonant tunneling of holes between the layers.

We model the interacting exciton system with the Hamiltonian $\begin{pmatrix} \mathcal{E}_{\text{iX}} & W_1 & W_2 \\ W_1^* & \mathcal{E}_1 & 0 \\ W_2^* & 0 & \mathcal{E}_2 \end{pmatrix}$. The diagonal elements describe the energy of uncoupled interlayer exciton $\mathcal{E}_{\text{iX}}$ and two intralayer moiré excitons $\mathcal{E}_1$ and $\mathcal{E}_2$ ($> \mathcal{E}_1$). The three-level model is sufficient because in experiment we observe no more than two intralayer moiré excitons with significant oscillator strength. The interlayer exciton energy is linear in vertical electric field, $\mathcal{E}_{\text{iX}} = \mathcal{E}_0 + D E_{appl}$, where $\mathcal{E}_0$ is the energy at zero field and $D$ is the out-of-plane effective electric dipole. The intralayer exciton energies are field-independent. The off-diagonal element $W_j$ and its complex conjugate $W_j^*$ are the coupling constant between the interlayer exciton and intralayer moiré exciton $X_j$ ($j$ = 1, 2). We assume the intralayer moiré excitons do not couple between themselves.

The black dashed lines in Fig. 2 are the three-level model that fits the experimental result best. In Fig. 2d-f we show a detailed comparison between the field-dependent exciton energies extracted from the experiment (symbols) and from the model with coupling (dashed lines) and without coupling (dotted lines). Comparison for the exciton oscillator strength is provided in Supplementary Fig. 2. The agreement between experiment and model is generally excellent. The extracted coupling constants $|W_1| \approx 40 - 50$ meV and $|W_2| \approx 20 - 30$ meV exceed the exciton linewidths ($\approx 10$ meV). The system is indeed in the strong coupling regime. The extracted effective electric dipole $D \approx 0.35 - 0.4$ $e$·nm ($e$ denoting the elementary charge) is also consistent with the value estimated from electrostatics $D \approx \frac{\epsilon_{hBN}}{\epsilon_W} et \approx 0.3$ $e$·nm. Here we take the layer separation $t \approx 0.7$ nm and the out-of-plane dielectric constant of hBN and TMDs $\epsilon_{hBN} \approx 3$ and $\epsilon_W \approx 7$ (see Methods for details).

The spatially mixed moiré excitons are also observed in a different TMD heterobilayer system – angle-aligned MoSe$_2$/WS$_2$ heterobilayers. Here the conduction bands of the two layers are closely aligned and hybridize. The spatially mixed excitons are the hybrid of the fundamental intralayer exciton in MoSe$_2$ and the interlayer exciton comprised of transitions from the upper valence band of MoSe$_2$ to the conduction band of WS$_2$ (Fig. 3a). Overall, the exciton coupling is substantially weaker in this system. But multiple spatially mixed moiré excitons can still be discerned in undoped MoSe$_2$/WS$_2$ (Fig. 3b). In particular, we identify two families of spatially mixed moiré excitons with distinct electric-field dispersions. The family around the second intralayer moiré exciton ($X_2$) has a larger effective electric dipole than the one around the third intralayer moiré exciton ($X_3$). The best fit to a two-level model yields $D$ ~ 0.24 and 0.17 $e$·nm for the $X_2$- and $X_3$-families, respectively (Fig. 3c). A plausible explanation for such a large difference in $D$ is interlayer exciton localization in different sites of the moiré superlattice formed by two incommensurate crystal lattices. This phenomenon is known for intralayer moiré excitons [12,17]. The effective electric dipole is sensitive to the local dielectric property and presumably can take different values at different moiré sites. The corresponding coupling constants for the $X_2$- and $X_3$-families are ~ 1 and 3 meV, respectively. They are smaller than the exciton linewidths (~ 5-10 meV). The system is not in the strong coupling regime and energy level anticrossing is not clearly resolved. The difference in coupling strengths between MoSe$_2$/WS$_2$ and WSe$_2$/WS$_2$ deserves further studies.

Finally we consider the doping effect on spatially mixed moiré excitons in WSe$_2$/WS$_2$ heterobilayers. Doping in MoSe$_2$/WS$_2$ heterobilayers further weakens the spatially mixed excitons and is not studied (see Supplementary Fig. 3). We focus on electron doping in WSe$_2$/WS$_2$, for which strongly interacting interlayer and intralayer excitons are observed. Figure 4 summarizes the fitting parameters of the three-level model as a function of filling factor $\nu$ (top axis) and gate voltage $V_G$ (bottom axis). At $V_G = 0$, the Fermi level is inside the energy gap of the heterobilayer. The vertical dashed lines correspond to $\nu = 0$, 1, and 2. The second and higher electron moiré minibands (for $\nu > 2$) are no longer well isolated. The most striking doping effect is shown in Fig. 4a for the moiré miniband edges of the uncoupled intralayer exciton ($\mathcal{E}_1$) and interlayer exciton at zero field ($\mathcal{E}_0$). Both $\mathcal{E}_1$ and $\mathcal{E}_0$ show plateaus around $v = 0$, 1, and 2 although $\mathcal{E}_1$ decreases and $\mathcal{E}_0$ increases with $v$. The order of the two states is altered at $1 < v < 2$. Figure 4b and 4c are the doping dependences of the coupling constants $|W_1|$ and $|W_2|$ and the effective dipole moment $D$, respectively.

Our result shows that the properties of spatially mixed moiré excitons can be effectively tuned by doping. And the electronic correlation effect induced by moiré superlattices is important to understand the experimental result. In the absence of strong correlation, for instance in angle-misaligned samples, we observe continuous exciton band renormalization upon doping (Fig. 1d). This is not compatible with Fig. 4a. Recent studies show the importance of on-site Coulomb repulsion between charge carriers on a moiré superlattice [3,11]. At $v = 1$, the strong on-site Coulomb repulsion suppresses the inter-site electron tunneling. It gives rise to a Mott insulating state with a large Coulomb gap between the lower and upper Hubbard bands. The Coulomb gap suppresses free carrier screening of the Coulomb interaction and the exciton band renormalization. Away from $v = 1$, the system becomes metallic again and effectively screens the Coulomb interaction. The energy plateau observed around $v = 1$ is thus likely related to the Mott insulating state. The interplay between free carrier screening of the Coulomb interaction and correlation-driven Coulomb gap can also explain the observed renormalization of the electric dipole of the interlayer excitons through that of the TMD dielectric constant. TMD moiré superlattices provide an exciting platform to understand, explore, and manipulate the spatially mixed moiré excitons and their interplay with emerging correlated states.

**Methods**
**Sample preparation and device fabrication**
The dual-gated WSe$_2$/WS$_2$ and MoSe$_2$/WS$_2$ hetero-bilayer devices were fabricated by the dry transfer method reported in the literature [3,32]. In short, atomically thin flakes were exfoliated from bulk crystals onto SiO$_2$/Si substrates. Angle aligned TMD flakes, gated and contacted by few-layer graphite flakes, were then stacked on top of each other. Angle alignment, with precision of about 0.5°, was achieved by angle-resolved optical second harmonic generation (SHG) measurements. Details of the SHG measurement have been reported by recent studies [3,12]. The SHG intensity from monolayer TMDs has a six-fold pattern, which correlates with the crystal axis and allows angle aligned sample transfer. Furthermore, the total SHG intensity is constructively enhanced and destructively suppressed for 0°- and 60°-samples, respectively, allowing us to distinguish between the two cases (Supplementary Fig. 4).

**Reflection contrast measurements**
Details of the reflection contrast measurement have been reported in ref. [3]. In short, white light with a continuous broadband spectrum from a halogen lamp was collimated and focused onto the sample with a diffraction-limited spot. The reflected light was collected by an objective and detected by a liquid-nitrogen cooled CCD camera coupled to a spectrometer. The reflectance contrast spectrum $R \equiv (I' - I)/I$ was obtained by comparing the reflected light intensity from the sample ($I'$) to a featureless spectrum ($I$) from the substrate. The sample is grounded during the optical measurements, and the top and bottom gate voltages were controlled by two sourcemeters (Keithley 2400). The device was mounted in a closed-cycle cryostat with temperature of 5 K (Montana Instruments).

**Calibration of doping density and electric field**
The doping density in the TMD bilayers was determined from the applied gate voltage using a parallel-plate capacitor model, details of which have been reported [3]. To summarize, the doping density was calculated by the sum of the two applied gate voltages and the geometric capacitances of the nearly symmetric top and bottom gates. The geometric capacitance is determined by the thickness of the hBN dielectric (measured by atomic force microscopy) and its dielectric constant $\epsilon_{hBN} \approx 3$ [33]. The filling factor was then determined from the doping density and the calculated moiré density based on the known lattice and alignment angle mismatch. Whereas symmetric gating induces doping, antisymmetric gating applies a vertical electric field to the sample. The electric field was calculated by the voltage difference applied to the two gates divided by the total thickness of the device measured by atomic force microscopy.

**Electrostatics model for interlayer excitons in TMD bilayers**
To obtain the effective electric dipole $D$ of interlayer excitons in TMD bilayers, we calculate the internal electric field $E_W$ within the bilayer for a given applied field $E_{appl}$, using a parallel-plate capacitor model. Following the steps provided by ref. [22], we obtain, for symmetric top and bottom gates,

$$E_W \approx \frac{\epsilon_{hBN}}{\epsilon_W} E_{appl} \qquad (1)$$

The hBN out-of-plane dielectric constant is $\epsilon_{hBN} \approx 3$ [33]. We approximate the out-of-plane dielectric constant of the TMD bilayer by the averaged value of the two layers, which is $\epsilon_W \approx 6.9$ for both WSe$_2$/WS$_2$ and MoSe$_2$/WS$_2$ [34]. For fixed doping density $n$, the Stark shift $\mathcal{E}_{iX}$ for interlayer excitons becomes

$$\mathcal{E}_{iX} = et \cdot E_W \approx et \frac{\epsilon_{hBN}}{\epsilon_W} E_{appl} = D E_{appl} \qquad (2)$$

The effective electric dipole is $D \approx \frac{\epsilon_{hBN}}{\epsilon_W} et$.

**Competing interests**
The authors declare no competing interests.

**Data availability**
The data that support the findings of this study are available within the paper and its Supplementary Information. Additional data are available from the corresponding authors upon request.

**Figures and figure captions**

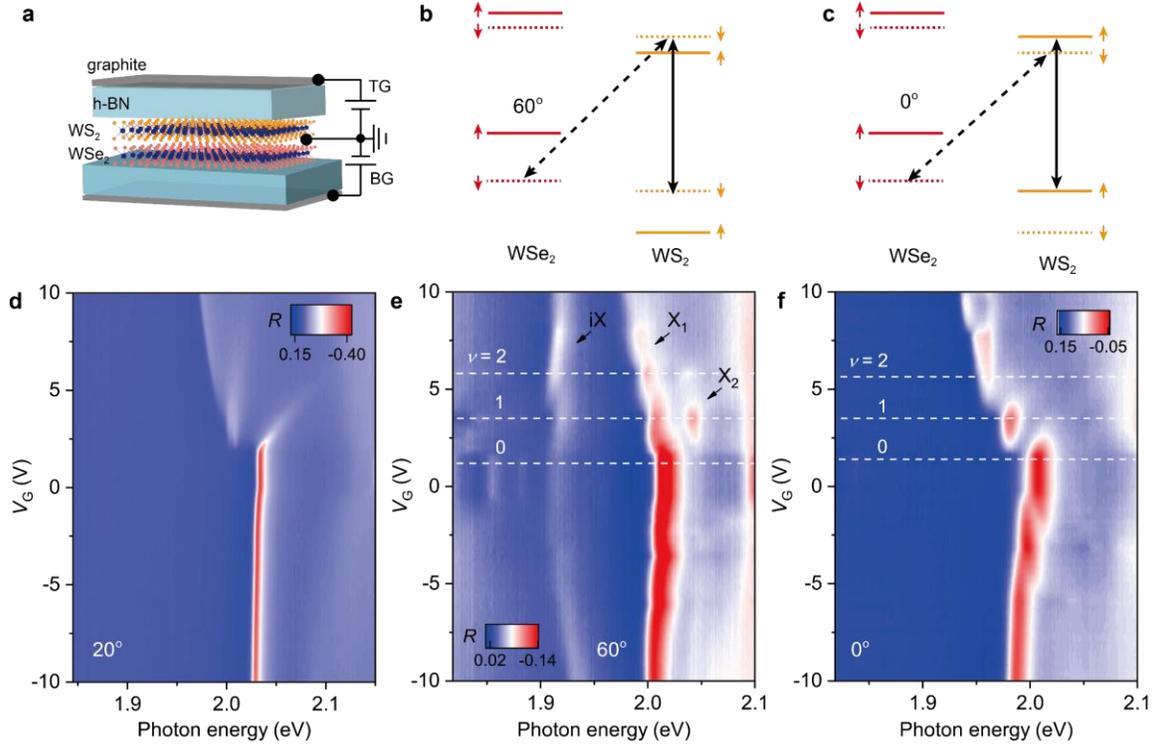

**Figure 1 | Doping dependent reflection contrast spectrum of WSe$_2$/WS$_2$ bilayer. a,** Schematic dual-gate device structure. The WSe$_2$/WS$_2$ bilayer is encapsulated by nearly symmetric top and bottom gates. **b, c,** Type-II band alignment for ~ 60° (**b**) and ~ 0° (**c**) aligned samples. Spin up (down) bands are denoted by solid (dotted) lines, and by single-sided arrows. Solid (dashed) double-sided arrows denote intralayer (interlayer) dipole-allowed optical transitions. Mixing of the two types of transitions is allowed only for the ~ 60° samples. **d, e, f,** Contour plot of the gate voltage dependent reflection contrast spectrum for ~ 20° (**d**), ~ 60° (**e**) and ~ 0° (**f**) samples. Gate voltage is applied symmetrically on both gates. The white dashed lines denote the gate voltages for zero, half, and full filling of the first moiré conduction band of the WS$_2$ layer. iX, X$_1$ and X$_2$ in **e** denote interlayer, first intralayer and second intralayer moiré excitons, respectively.

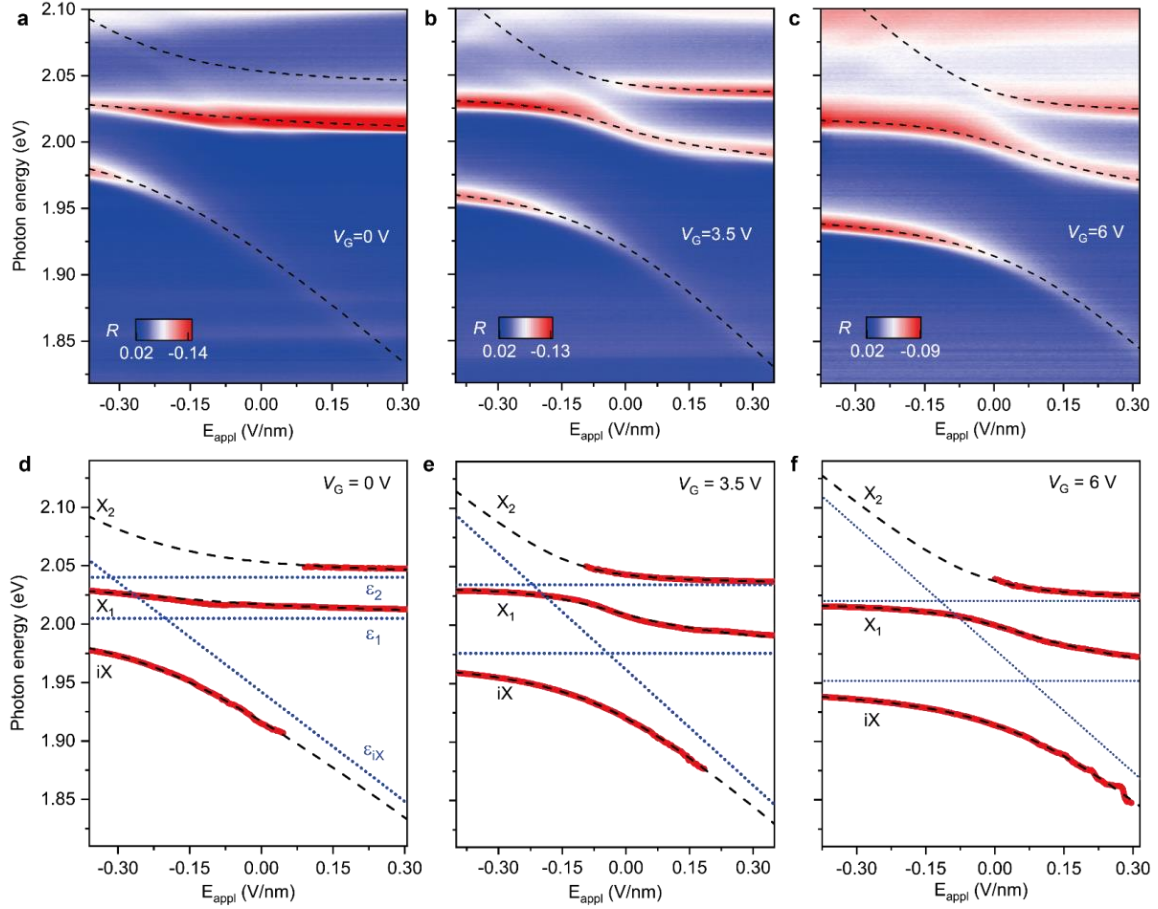

**Figure 2 | Electric field dependence of spatially mixed exciton resonances for WSe$_2$/WS$_2$ bilayer. a**, **b**, **c,** Contour plot of the electric field dependent reflection contrast spectrum for the ~ 60° sample at different doping densities. Electric field is applied by varying the voltage difference between the two symmetric gates. The sum of the gate voltage is fixed at each panel so that a constant doping density is maintained. The dashed lines denote the theoretical fittings of the hybridized exciton resonances. **d, e, f,** Extracted electric field dependence of the exciton resonance energies at doping densities corresponding to **a**, **b, c**. Dotted lines represent the dispersion for the uncoupled exciton states. Dashed lines denote the dispersion for the hybridized exciton states obtained by fitting the experimental results to the theoretical model presented in the main text.

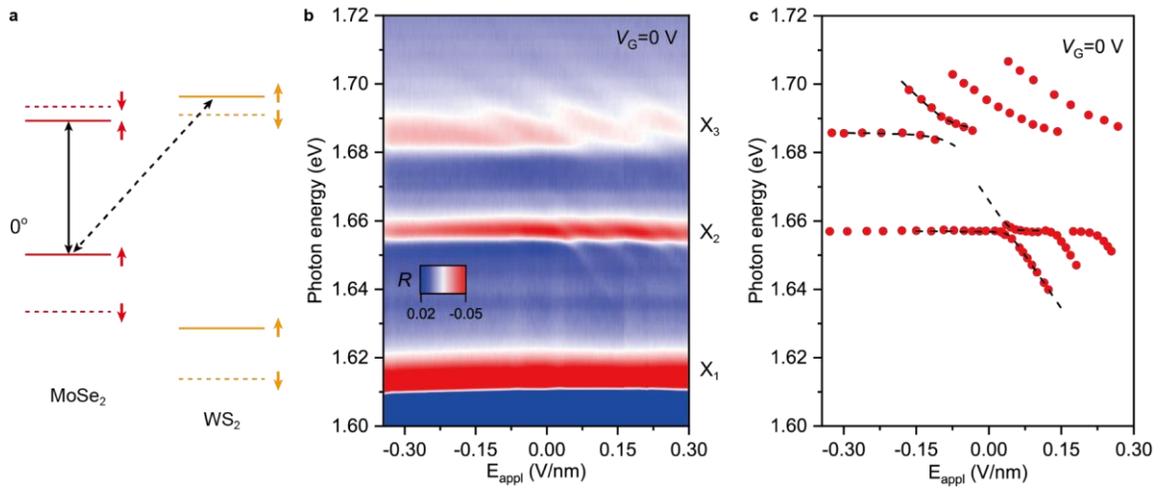

**Figure 3 | Electric field dependence of spatially mixed exciton resonances for MoSe$_2$/WS$_2$ bilayer. a,** Type-II band alignment for ~ 0° aligned samples. Mixing of intralayer and interlayer excitons is allowed for both ~ 0° and ~ 60° samples because of the small spin-orbit splitting for the conduction bands. **b,** Contour plot of the electric field dependent reflection contrast spectrum at zero gate voltage. **c,** Extracted electric field dependence of the exciton resonance energies. Dashed lines are theoretical fits from a two-level model (see main text). Multiple moiré exciton species are seen in MoSe$_2$/WS$_2$.

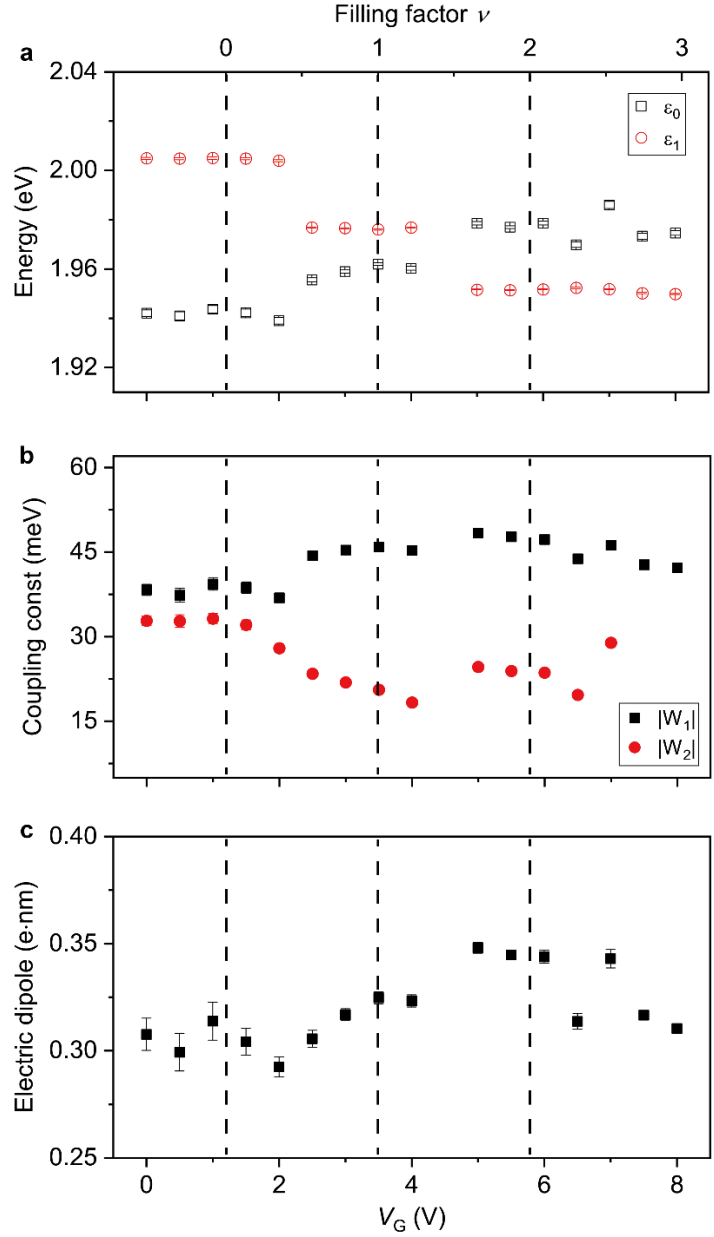

**Figure 4 | Filling factor dependence. a, b, c,** Filling factor dependence of uncoupled intralayer and interlayer exciton resonance energies at zero electric field (**a**), intralayer-interlayer exciton coupling strengths (**b**), and interlayer exciton effective electric dipole (**c**). Vertical dashed lines denote zero, half and full fillings of the first moiré conduction band of the $WS_2$ layer.

**Supplementary figures and figure captions**

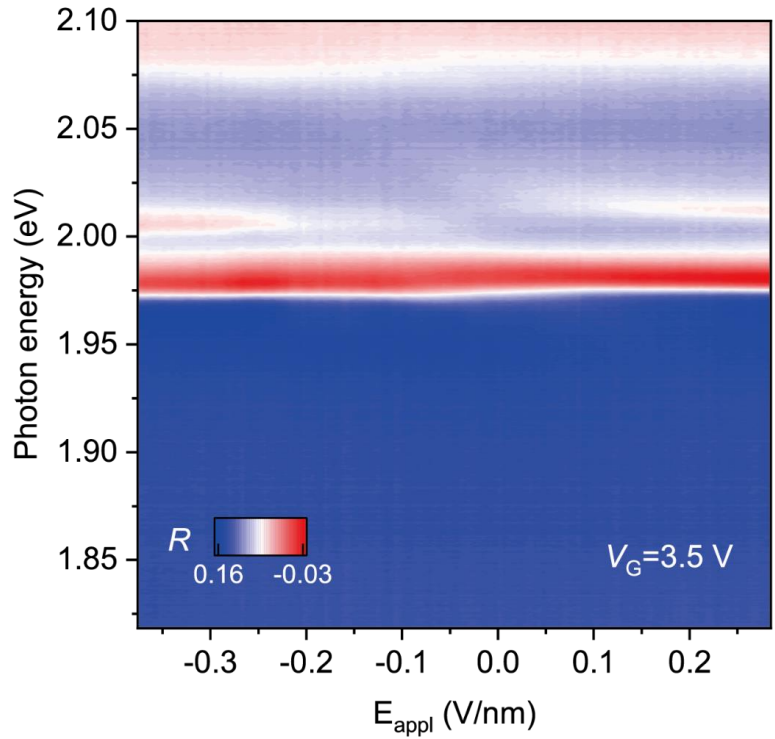

**Supplementary Figure 1 | Electric field dependent reflection contrast spectrum of 0° aligned WSe$_2$/WS$_2$ bilayer.** Contour plot of the electric field dependent reflection contrast spectrum for a ~ 0° sample at fixed doping density. No interlayer exciton feature and hybridization of exciton states is seen. Only non-dispersive intralayer exciton resonances from the WS$_2$ layer are observed.

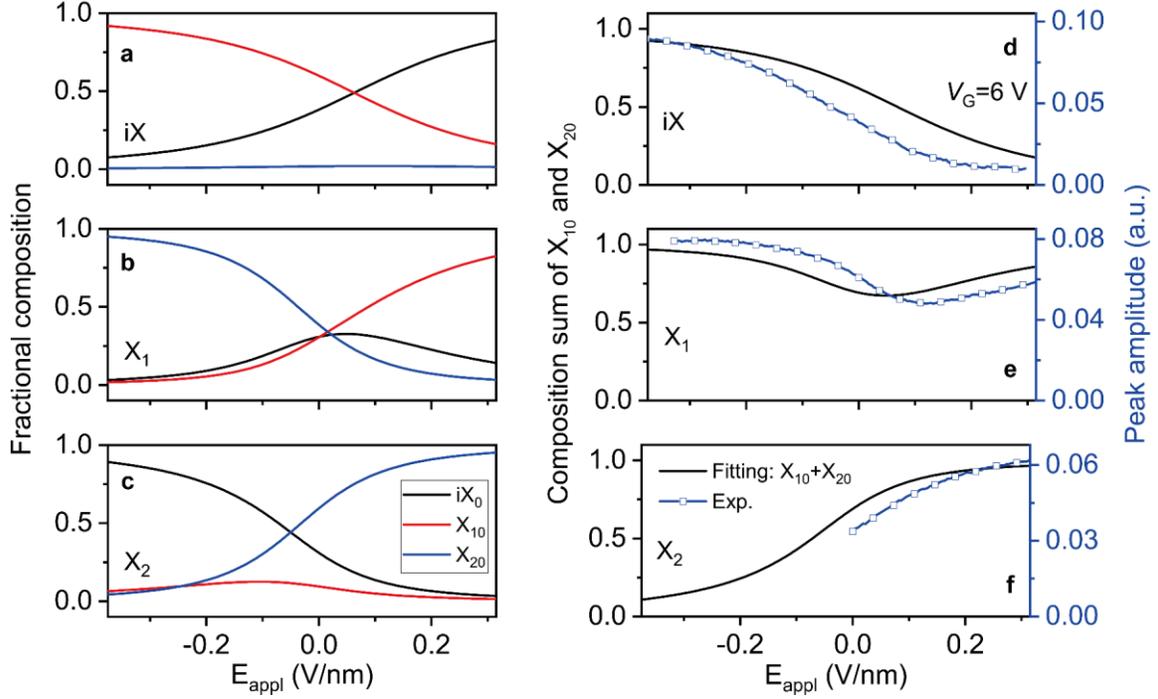

**Supplementary Figure 2 | Electric field dependence of mixed exciton amplitudes. a, b, c,** Decomposition of the amplitude squared of the uncoupled exciton states for the hybridized iX (**a**), $X_1$ (**b**) and $X_2$ (**c**) states as a function of electric field. Black, red and blue lines denote the amplitude squared of the uncoupled interlayer exciton $iX_0$, intralayer exciton $X_{10}$ and intralayer exciton $X_{20}$ states, respectively. The results are obtained by solving the eigenvalue equation for the three-level system presented in the main text. All of the states in the three-level Hamiltonian are normalized. **d, e, f,** Electric field dependence of the sum of the uncoupled $X_{10}$ and $X_{20}$ amplitude squared (solid line) for the hybridized iX (**d**), $X_1$ (**e**) and $X_2$ (**f**) states. The calculated results are compared to the experimental electric field dependence of the peak amplitude (proportional to oscillator strength) of the hybridized iX (**d**), $X_1$ (**e**) and $X_2$ (**f**) excitons (symbols connected by blue lines). Since only the uncoupled intralayer excitons $X_{10}$ and $X_{20}$ couple to light strongly, the oscillator strengths of the hybridized excitons are well approximated by the sum of these two contributions. The experimental results agree well with the theoretical model.

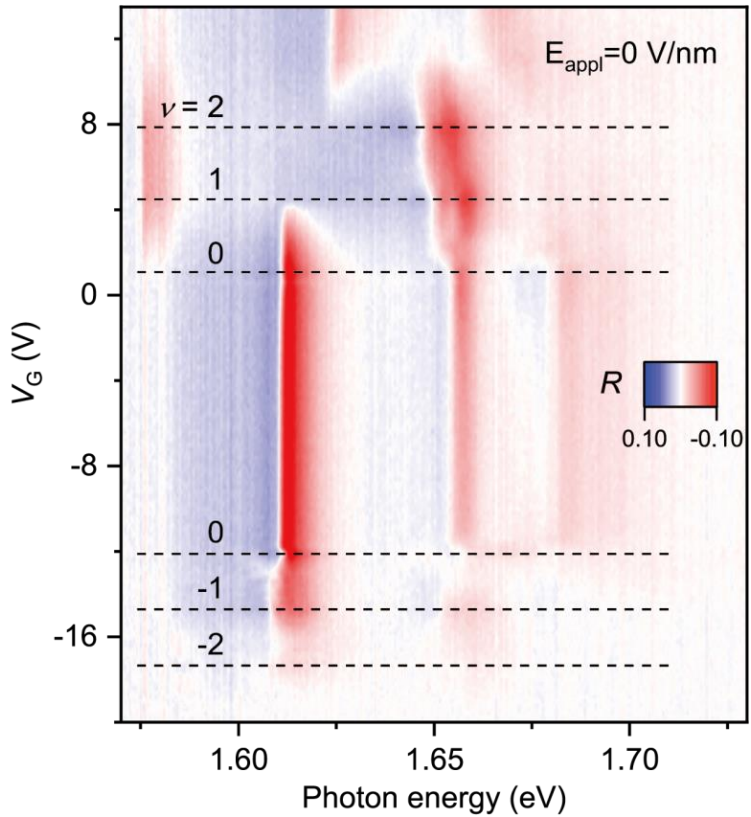

**Supplementary Figure 3 | Doping dependent reflection contrast spectrum of angle aligned MoSe$_2$/WS$_2$ bilayer.** Contour plot of the gate voltage dependent reflection contrast spectrum for a ~ 0° aligned MoSe$_2$/WS$_2$ sample. The dashed lines denote the gate voltages for zero, half, and full filling of the first moiré conduction (positive gate voltage) and valence (negative gate voltage) bands. The oscillator strengths of the moiré excitons are significantly reduced with doping.

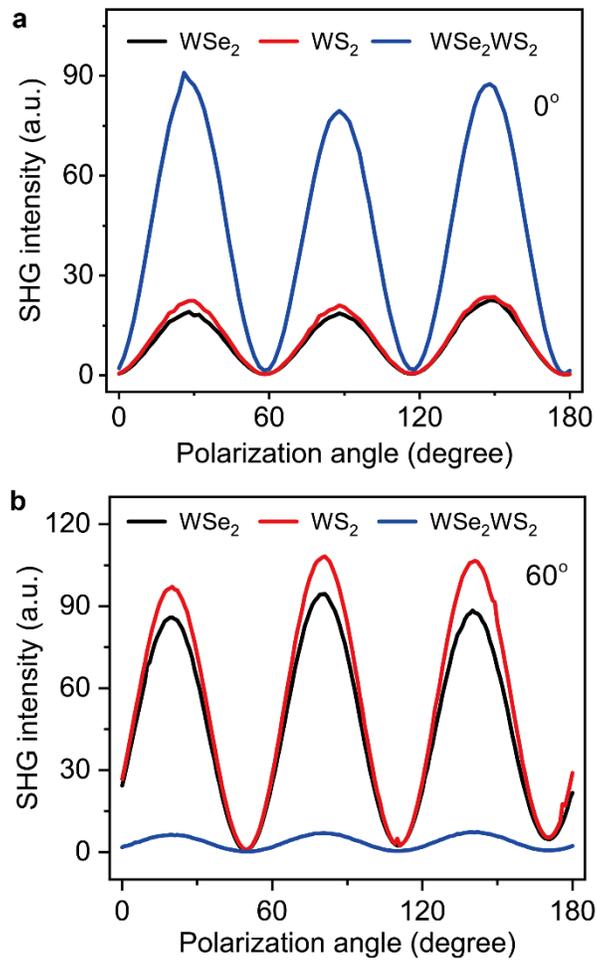

**Supplementary Figure 4 | Polarization resolved SHG intensity. a, b,** Dependence of the SHG intensity on the incident polarization angle for isolated $WSe_2$ monolayer (black), isolated $WS_2$ monolayer (red) and the coupled $WSe_2/WS_2$ bilayer with ~ 0° (**a**) and ~ 60° (**b**) alignment angle. Only the cross-polarized SHG component is detected. Whereas the ~ 0° bilayer has substantially stronger SHG intensity than the isolated monolayers due to constructive interference, the reverse happens for the ~ 60° bilayer due to destructive interference. The result allows us to clearly distinguish the two cases.